\documentclass[prb,twocolumn,superscriptaddress]{revtex4}
\usepackage{graphicx}
\usepackage{amssymb}
\usepackage{amsmath}
\usepackage{xspace}

\begin{document}
\title{Localized polarons and doorway vibrons in finite quantum structures}
\author{H.~Fehske}
\affiliation{
Institut f\"ur Physik, Ernst-Moritz-Arndt-Universit{\"a}t
Greifswald, 17487 Greifswald, Germany }
\author{G. Wellein}
\affiliation{
Regionales Rechenzentrum Erlangen, Universit\"at Erlangen-N\"urnberg,  
91058 Erlangen, Germany }
\author{J. Loos}
\affiliation{Institute of Physics, Academy of 
Sciences of the Czech Republic, 16200 Prague}
\author{A. R. Bishop}
{\affiliation{\mbox{Theory, Simulation and Computation Directorate, 
Los Alamos National Laboratory, Los Alamos, New Mexico 87545}}
\begin{abstract}
We consider transport through finite quantum systems such as
quantum barriers, wells, dots or junctions, 
coupled to local vibrational modes in the quantal regime. 
As a generic model we study the Holstein-Hubbard 
Hamiltonian with site-dependent potentials and interactions.   
Depending on the barrier height to electron-phonon coupling strength
ratio and the phonon frequency we find distinct opposed behaviors: 
Vibration-mediated tunneling or intrinsic localization of (bi)polarons. 
These regimes are strongly manifested in the density correlations, 
mobility, and optical response calculated by exact numerical techniques. 
\end{abstract}
\maketitle
\section{Introduction}
Recent progress in nanotechnology has triggered a systematic
study of electronic transport in microscopic systems weakly coupled
to external electrodes~\cite{Nano}. In such devices the active element 
can be a single organic molecule, but also a suspended Carbon 
nanotube, and may be thought of as a quantum dot contacted  
to metallic leads that act as macroscopic charge reservoirs. 
In small quantum dots energy level quantization becomes 
as important as electron correlations. Additionally 
vibrational modes play a central role in the electron
transfer through quantum dots or molecular junctions
(see, e.g., the topical review Ref.~\onlinecite{GRN07}). 

The electron-phonon (EP) interaction is found to particularly 
affect the dot-lead 
coupling. Here electronic and vibrational energies can become 
of the same order of magnitude, e.g. when Coulomb charging is reduced 
by screening due to the electrodes~\cite{NCUB07}. The same circumstance
holds in the polaron crossover regime, where the electrons are 
dressed by a phonon cloud, implying that
phonon features for the current through the quantum device 
are of major importance~\cite{ZM07}.  
Phonon and polaron effects in nanoscale devices have been 
extensively discussed e.g. for (magnetic) molecular 
transistors~\cite{MAM04,CGN05,NCUB07},
quantum dots~\cite{qdot}, tunneling diodes and Aharonov-Bohm rings~\cite{BT95}, 
metal/organic/metal structures~\cite{YSSB99}, or 
Carbon nanotubes~\cite{LLKD04}.

In this paper we study the electronic properties of   
various EP coupled quantum systems.  We consider  
one-dimensional  structures, where the ``quantum device''
is sandwiched between two metallic wires characterized 
by (tight-binding) electron hopping amplitude $t$, local 
Coulomb interaction $U$, and EP coupling $\varepsilon_p$ 
(cf. Fig.~\ref{Fig1}). Such systems may be described by a 
generalized Holstein-Hubbard Hamiltonian. 
The Holstein-Hubbard model~\cite{FILTB94,BKT00,FWHWB04,HAL05} is not 
completely realistic, of course, as it only includes
local electron-phonon and electron-electron interactions
as well as a coupling to (dispersionless) optical phonons.
However, we are interested in fundamental phenomena arising
from the combination of electron-phonon interaction
and ``confinement'' in discrete quantum structures.
Besides many aspects of finite (EP coupled) quantum systems 
may be understood using such simplified effective 
models~\cite{MAM04,CGN05,JHW98}.  
\begin{figure}[b]
  \includegraphics[width=\linewidth]{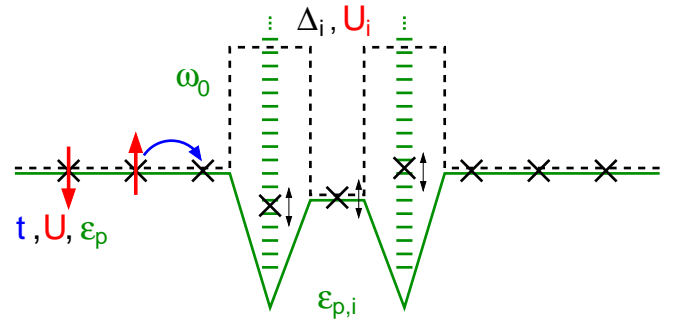}
\caption{(Color online) Schematic representation of model devices  
described by the Hamiltonian~\eqref{hhm}.}
\label{Fig1}
\end{figure}
\section{Model}
Allowing for site-dependent potentials and electron-phonon/electron 
interactions, the tight-binding Holstein-Hubbard Hamiltonian takes the form 
\begin{eqnarray}
\label{hhm}
H &=& \sum_{i,\sigma}\bar{\Delta}_i n_{i\sigma} - t\sum_{i,\sigma}
(c^{\dagger}_{i\sigma}c^{}_{i+1 \sigma}
+\mbox{H.c.})+\omega_0\sum_{i} b_i^{\dagger} b_i^{}\nonumber\\
&&-\sum_{i,\sigma}\bar{g}_i\omega_0\,
(b_i^{\dagger}+b_i^{})n_{i\sigma} +\sum_i 
\bar{U}_i n_{i\uparrow}n_{i\downarrow}\,.
\end{eqnarray}
Here $\bar{\Delta}_i=\Delta+\Delta_i$, 
where the potentials, $\Delta_i$ on site $i$, 
can describe a tunnel barrier, disorder, or a voltage basis. 
Since we will treat left and right leads in equilibrium,  
we choose $\Delta=0$ throughout the sample,
neglecting a bias between the metal leads and, in order to avoid
spurious multi-scattering from the boundaries in a finite system, 
we take periodic boundary conditions. The parameter  
$U_i$ ($\bar{U}_i=U+U_i$), can be viewed as additional 
Hubbard interaction or charging energy of, e.g., a quantum dot 
molecule. The parameter  
$\bar{g}_i=[(\varepsilon_{p}+\varepsilon_{p,i})/ \omega_0]^{1/2}$
describes the local coupling of an electron on site $i$ to  
an internal optical vibrational mode at the 
same site~\cite{phononcouplingcomment}. 
Here $(\varepsilon_{p}+\varepsilon_{p,i})$
denotes the corresponding polaron binding energy, and 
$\omega_0$ is the frequency of the optical 
phonon~\cite{phononfrequencycomment}. 
In this way the model, e.g., mimics tunneling through 
(single or double) barriers ($\Delta_i>0$), 
trapping of electrons, polarons, or bipolarons 
at single-impurity or double-well sites ($\Delta_i<0$), 
or transport through quantum dots with soft dot-lead links.

On a translational invariant lattice ($\bar{\Delta}_i=\Delta$,
$\bar{g}_i=g$, $\bar{U}_i=U$) the Holstein-Hubbard model  
can be numerically solved by variational diagonalization 
in the one- and two-particle sectors of interest here. This holds 
in the thermodynamic limit, for the whole range of parameters and 
any dimension (for a recent review of the Holstein (bi)polaron problem 
see Ref.~\onlinecite{FT07}). 
The main result is a {\sl continuous cross over} with
increasing EP coupling strength, from 
electronic quasiparticles weakly renormalized by phonons 
to (small) polarons or bipolarons~\cite{WRF96}. 
Depending on the value of the adiabaticity ratio
$\alpha=\omega_0/t$, in one-dimensional systems,  
the large-to-small polaron cross over 
is determined by the more restrictive of the two conditions
$\lambda=\varepsilon_{p}/2t\geq 1$ (relevant for $\alpha\ll 1$,
adiabatic regime) or $g^2\geq 1$ (for $\alpha\gg 1$,
anti-adiabatic regime)~\cite{CSG97}.

Here we address the problems of polaron/bipolaron formation
and phonon-assisted transport for the more complicated {\it inhomogeneous}
barrier structures and interactions described by the above Hamiltonian. 

\section{Numerical Results and Discussion}

In our numerical work we combine exact diagonalization (ED) and 
kernel polynomial methods~\cite{WWAF06,JF07} to determine the
ground-state and spectral properties.  
All energies will be measured in units of $t$.

\subsection{Single-electron case}
We first consider a single electron that 
tunnels through a {\it single quantum barrier}. 
The barrier height is assumed to considerably exceed
the  electron half-bandwidth. Outside the barrier 
the electron is subjected to a rather moderate EP 
coupling, $\varepsilon_{p}=0.5$. The chosen phonon frequency
$\omega_{0}=0.4$ reflects an adiabatic situation.

Figure~\ref{Fig2} shows the behavior of the system's kinetic energy
\begin{equation}
E_{kin}= - \sum_{i,\sigma} \langle(c^{\dagger}_{i\sigma}c^{}_{i+1 \sigma}
+\mbox{H.c.})\rangle
\end{equation} 
as the EP coupling strength 
is increased at the barrier site.
Recall that both coherent and incoherent transport processes contribute 
to $E_{kin}$. Without loss of generality we assume the barrier 
to be located at site $4$.  For $\varepsilon_{p,4}=0$ the barrier is 
almost impermeable, 
consequently the local electron density $n_{e,i}=\langle n_{i\uparrow}
+n_{i\downarrow}\rangle$ is near zero at site $4$. 
An additional local EP interaction $\varepsilon_{p,4}$ 
renormalizes the on-site adiabatic potential, i.e., it leads to
a local polaronic level shift that softens the barrier.
Note that the kinetic energy stays almost constant
until $\varepsilon_{p,4}$ exceeds a certain critical
value, $\varepsilon_{p,4}^c$. At $\varepsilon_{p,4}^c$, 
the mobility of the electron is arrested, and the charge carrier 
becomes quasi localized at the barrier site~\cite{estimate}. 
The large number of bound vibrational states ($n_{ph,4}=\langle b_4^\dagger 
b_4\rangle\simeq 10$) give rise to a displaced oscillator state 
at site $4$, i.e. a new equilibrium state of the lattice
results, which lowers the energy.
The {\it jump-like transition} is in striking contrast to what is observed if 
we increase only the EP coupling locally (without having a barrier), 
or if we form a quantum well ($\Delta_4<0$) without additional 
EP interaction (see inset of Fig.~\ref{Fig2}). In these cases 
we found a gradual transition from a nearly free electron to
a rather immobile particle. 
\begin{figure}[t]
  \includegraphics[width=\linewidth]{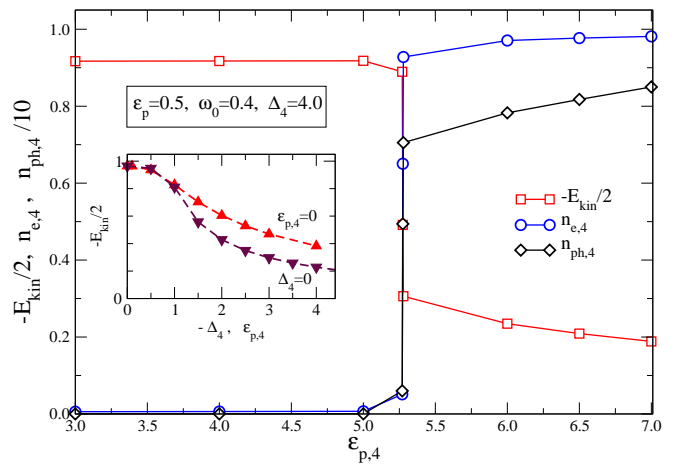}
\caption{(Color online) Kinetic energy of a single electron on a $N=8$ site
ring with potential barrier $\Delta_4=4$ at site $i=4$.
The main panel gives $E_{kin}$ (squares), the electron density 
$n_{e,4}$ (circles), and the mean phonon number $n_{ph,4}$ (diamonds) 
at the barrier site  
as functions of an additional EP coupling  $\varepsilon_{p,4}$. 
The inset shows the variation of $E_{kin}$ if the 
potential $\Delta_4$ is lowered keeping  $\varepsilon_{p,4}=0$ (triangles up),
or if $\varepsilon_{p,4}$ is raised with  $\Delta_4=0$ (triangles down).} 
\label{Fig2}
\end{figure}

The extremely sharp polaron transition is accompanied by a 
drastic change in the optical response. The regular part 
of the optical conductivity is given by 
\begin{equation}
\sigma_{reg}(\omega) = 
   \sum_{n>0} \frac{|\langle n | \hat{j} |0\rangle |^2}{\omega_n}
   \delta(\omega - \omega_n)\,,
\label{sigmareg}
\end{equation}
where $\hat{j}=\text{i}et\sum_{i,\sigma}(c^{\dagger}_{i\sigma}c^{}_{i+1 \sigma}-
c^{\dagger}_{i+1\sigma}c^{}_{i \sigma})$ is the current operator and $|n\rangle$ 
label the eigenstates of $H$ with excitation energy $\omega_n=E_n-E_0$.

Figure~\ref{Fig3} shows $\sigma_{reg}(\omega)$ and the 
integrated spectral weight
\begin{equation}
S_{reg}(\omega)=\int_0^\omega\sigma_{reg}(\omega^\prime)d\omega^\prime\,,
\end{equation} 
in the vicinity of 
the transition, where a tiny increase of  
$\varepsilon_{p,4}$ (of about $8\times 10^{-3}$, 
from top to bottom) substantially changes the optical spectra.
While the upper panel resembles the optical spectra of
a large polaron with an absorption maximum at small
frequencies and a rather asymmetric line shape, we found a bimodal
signature near the transition point (middle panel) and finally the typical
(almost symmetric) small polaron absorption  just above 
$\varepsilon_{p,4}^c$ (lower panel). In this manner the system 
acts as an optical switch.

Corresponding behavior is found if we increase the barrier (voltage bias) 
keeping $\varepsilon_{p,4}$ fixed (see Fig.~\ref{Fig4}). Again 
the transition is ``discontinuous'' for small phonon frequencies, where the 
concept of an adiabatic energy surface holds to a good approximation.
At larger phonon frequencies non-adiabatic effects become increasingly 
important. Here the EP coupling does not work against the (static) barrier  
directly and the transition softens as in normal polaronic systems.
Furthermore, for $\omega_0\gg 1$, the EP coupling constant 
$\bar{g}_4$ is reduced (i.e., although $\varepsilon_{p,4}$ is fixed
we leave the strong coupling regime). 

\begin{figure}
  \includegraphics[width=\linewidth]{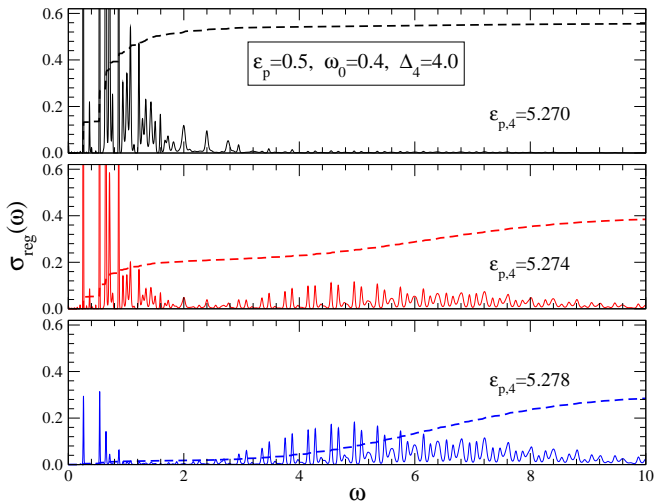}
\caption{(Color online) Optical response for the single-barrier 
system for various
$\varepsilon_{p,4}$. Dashed lines give the integrated weight $S_{reg}(\omega)$.}
\label{Fig3}
\end{figure}

In Fig.~\ref{Fig4} we have included the results obtained by a simple 
approximative analytical approach to the single-barrier problem. 
Assume that 
\begin{equation}
p_j=a(R) \;e^{|i-j|/R}
\end{equation}
 with $a(R)=\tanh \frac{1}{2R}$ 
is the probability for finding the particle at site $j$ 
away from the barrier site $i$.
Then, for the infinite system, the ground-state energy of a polaron 
with radius $R$, where $R=\infty$ corresponds to the free electron
while $R=0$ describes a small polaron localized at the impurity site, 
is given as 
\begin{equation}
E(R)=E_{loc}^{va}+E_{kin}^{va}
\end{equation}
with
\begin{align}
\label{evar}
E_{loc}^{va}&=\Delta_i a(R) - 
\omega_0\bar{g}^2_ia^2(R) [2-a(R)]\nonumber\\
&-2\omega_0 g^2 a^2(R)\sum_{i\geq 1}^N \text{e}^{-\frac{2i}{R}} 
\left[2-\text{e}^{-\frac{i}{R}}a(R)\right]\,,\\
E_{kin}^{va}&=-4t \text{e}^{-\frac{1}{2R}}a(R)\Bigg[
\exp \left\{-\frac{1}{2} a^2(R) \left[\bar{g}^2_i+g^2
\text{e}^{-\frac{2}{R}}\right]\right\}\nonumber\\
&+\sum_{i\geq 1}^N 
\text{e}^{-\frac{i}{R}}\,
\exp\left\{-\frac{1}{2} g^2 a^2(R) 
\text{e}^{-\frac{2i}{R}}\left[1+ \text{e}^{-\frac{2}{R}}\right]\right\}\Bigg]\,. 
\end{align} 
Of course, $E(R)$ has to be minimized with respect to $R$. 
Although the kinetic energy calculated in this way 
neglects important contributions from multi-phonon 
processes~\cite{commentekin}, we see that $E_{kin}^{va}$ gives 
a reasonable estimate for the critical value of $\Delta_4^c$,
at least in the adiabatic regime. In the anti-adiabatic region, 
$E_{kin}^{va}$ fails to describe the observed continuous cross over.
This is a well-known shortcoming of such a kind variational
approaches, which normally yield an abrupt polaron transition
in the whole frequency range~\cite{FILTB94}.

\begin{figure}
  \includegraphics[width=\linewidth]{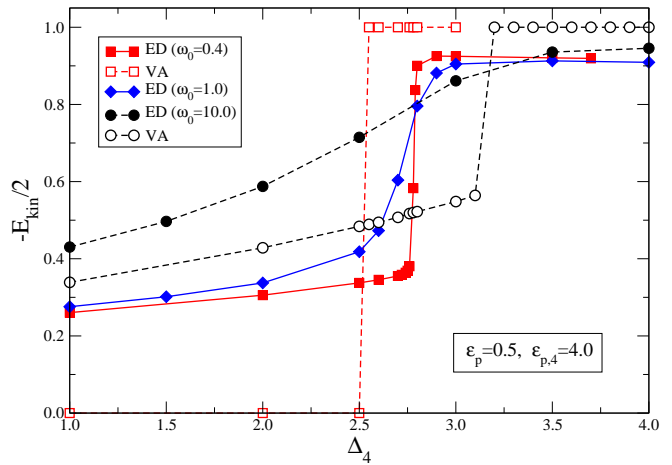}
\caption{(Color online) Kinetic energy upon increasing the barrier 
height $\Delta_{4}$
at constant EP coupling. Filled (open) symbols denote ED (analytical) 
results for $N=8$ ($N=\infty$).}
\label{Fig4}
\end{figure}

\subsection{Two-electron case}

Next we investigate two electrons in a single-barrier structure. 
Now, increasing the EP coupling on a barrier site with strong 
Coulomb repulsion, we found two successive transitions, see Fig.~\ref{Fig5}. 
In the first step one electron becomes localized at the barrier site 
blocking, because of the large $U$, the second one. 
Raising $\varepsilon_{p,4}$ further, both particles will be trapped, 
forming an on-site bipolaron. This can be seen most clearly
by monitoring the density correlation 
\begin{equation}
d_{i}= \tfrac{1}{4} ( n_{e,i} + 2 \langle n_{i\uparrow}n_{i\downarrow}\rangle )
\end{equation}
as a function of the EP coupling.
$E_{kin}$ and $n_{ph,4}$ clearly also show this two-step transition,
being related to significant changes of the ground-state 
phonon distribution~\cite{WRF96,JF07}, $|c_m|^2$, see insets.
The comparison of data for $N=8$, 10 shows 
that there is almost no finite-size dependence 
of the results.   
\begin{figure}
  \includegraphics[width=\linewidth]{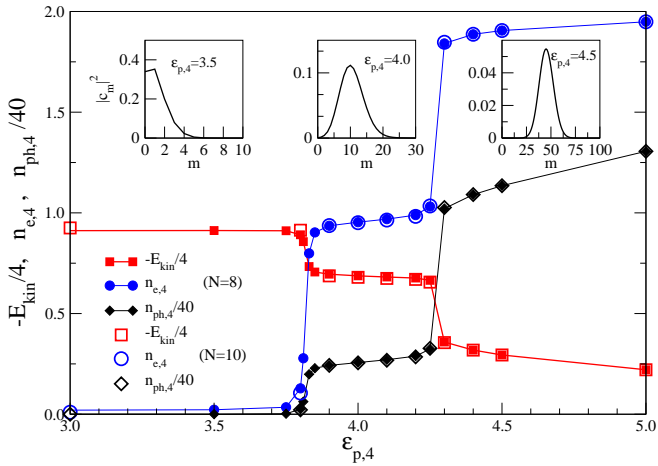}
\caption{(Color online) 
Kinetic energy (squares) and mean electron/phonon numbers
(circles/diamonds) for the case of two electrons in a singlet state.
We assumed strong Hubbard interaction at the barrier ($U_4=10$, $\Delta_4=2.5$).
Otherwise $U=0$, $\varepsilon_{p}=0.5$, and $\omega_0=0.4$.  
The insets display the weight of the $m$-phonon state in the 
ground state, $|c_m|^2$, for several characteristic $\varepsilon_{p,4}$.}
\label{Fig5}
\end{figure}

Finally let us consider the {\it double-barrier quantum dot structure}
sketched in Fig.~\ref{Fig1}, with two electrons in the system.
\begin{figure}
  \includegraphics[width=\linewidth]{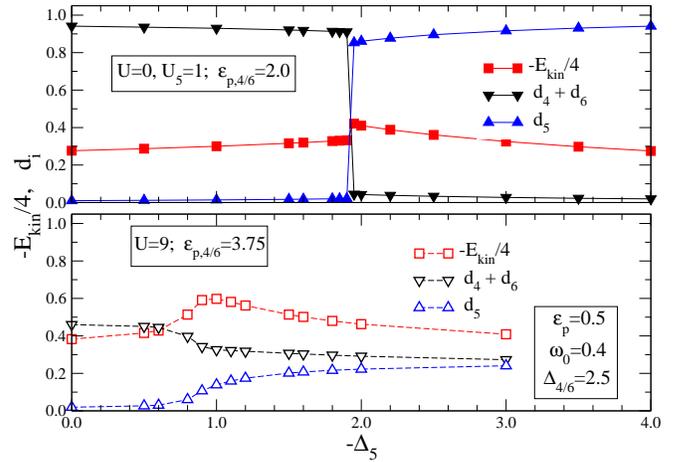}
\caption{(Color online) Two electrons in a quantum double-barrier  system
($\Delta_{4,6}=2.5$), where 
the potential of the central site $\Delta_{5}$ is lowered 
with respect to the leads (cf. Fig.~\ref{Fig1}).
Shown are the kinetic energy (squares) and the on-site
density correlations (triangles) for $\varepsilon_p=0.5$, 
$\omega_0=0.4$ and $U=0$, $U_5=1$
in the upper panel, while $U=9$, $U_5=0$ in the lower panel.}
\label{Fig6}
\end{figure}
We plot in Fig.~\ref{Fig6} the kinetic energy and the particle
occupation of the barrier and embedded dot 
sites as functions of the depth of the quantum well ($\Delta_5<0$). 
The upper panel describes the regime of moderate Coulomb
interaction at the dot, with $U=0$ otherwise. Here the dot
is unoccupied until its potential is lowered below
a critical value. Then the particles initially located
together at one of the dot-lead sites are transferred onto the
dot. In this process they change their nature
from a bipolaronic quasiparticle to two electrons  
solely (linearly dependent) bound by the potential well (impurity). 
Thus the ground state is a multi-phonon (few-phonon) state
for $\Delta_5>\Delta_5^c$ ($\Delta_5<\Delta_5^c$).  
If the system has a large Coulomb interaction everywhere, 
double occupancy is prohibited (lower panel). Then we find
initially one polaron per barrier (lead-dot) site and only
one particle tunnels to the dot at $\Delta_5^c$, thereby
stripping its phonons away. Note that the mobility is 
enhanced in the transition region.
\begin{figure}
  \includegraphics[width=\linewidth]{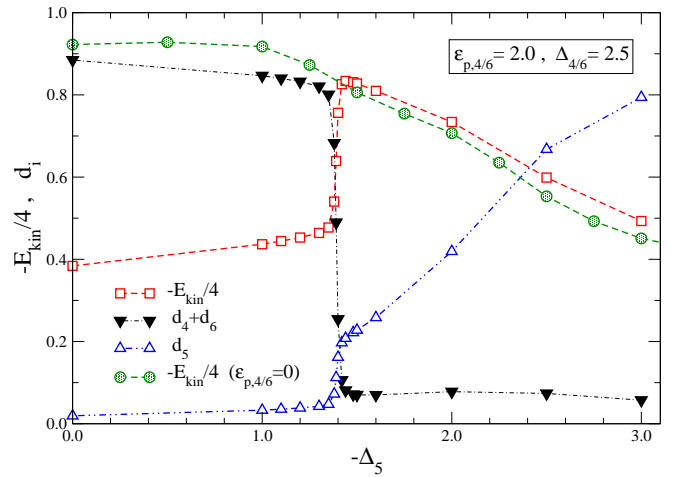}
\caption{(Color online) Quantum-well configuration with two electrons. 
Parameters
as in Fig.~\ref{Fig6} (upper panel) but now $\varepsilon_{p}=0$. 
For comparison we show $E_{kin}$ for a reference system 
without any EP coupling at the links.} 
\label{Fig7}
\end{figure}
\begin{figure}
  \includegraphics[width=\linewidth]{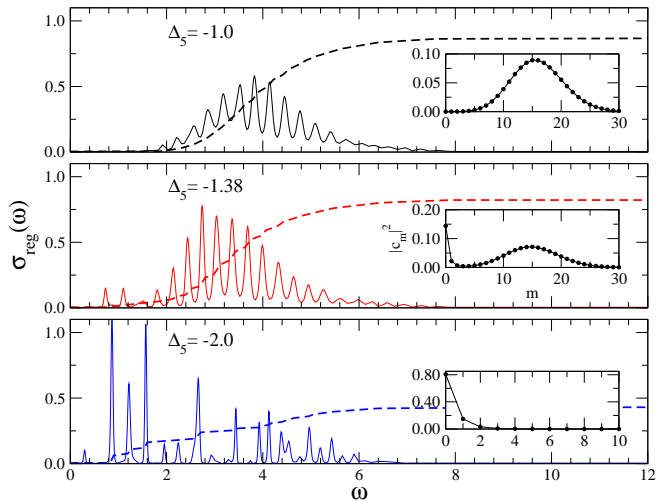}
\caption{(Color online) Optical conductivity for a ``soft-linked'' quantum dot
system with two electrons (parameters as in Fig.~\ref{Fig7}). 
The insets give the corresponding phonon distribution functions.
}
\label{Fig8}
\end{figure}

This effect is even more pronounced if we suppose  that the EP coupling 
acts on the dot-lead link sites only. As can be seen from
Fig.~\ref{Fig7} there is large jump-like increase of the particle's  
kinetic energy if the quantum well reaches $\Delta_5^c$. 
At $\Delta_5^c$ the bipolaron, located at one of the dot-lead link sites, 
dissolves and the electrons can pass over to the dot. 
Clearly $E_{kin}$ decreases if we lower the potential of the quantum 
dot further but note that for $\Delta_5<\Delta_5^c$ the kinetic 
energy is still {\it larger} than for a reference system 
without EP coupled  dot-lead link sites. 
In this way the local coupling to vibrational degrees of freedom of 
the barrier opens the gate for particle transmission, i.e.,  
vibronic excitations play the role of ``doorway states''.

To corroborate the importance of these quantum lattice fluctuation 
effects we determined the optical spectra below, 
near, and above the threshold $\Delta_5^c$. 
The data presented in Fig.~\ref{Fig8} 
give clear evidence for (bi)polaron hopping transport 
for a shallow quantum well, with dominant phonon emission 
and absorption processes, but  resonant  
vibration-mediated tunneling takes place for a deeper well. 

We emphasize that the increase of $E_{kin}$ in passing below 
$\Delta_5^c$ is accompanied by a decrease of the total integrated 
weight $S_{reg}(\infty)$ of the regular (incoherent) part of $\sigma_{reg}(\omega)$   
(compare dashed lines in Fig.~\ref{Fig8} from top to bottom).  
Thus, exploiting the f-sum rule, 
\begin{equation}
-E_{kin}/2=D+S_{reg}(\infty)\,,
\end{equation}
we can conclude that the
coherent contribution (Drude part $D$) to $E_{kin}$ is amplified.
The insets substantiate this interpretation. Starting from 
a Poisson-like distribution of $|c_m|^2$, a second maximum
develops at $m=0$ for $\Delta_5\gtrsim \Delta_5^c$, and finally,
for $\Delta_5<\Delta_5^c$, the ground state contains only 
zero-, one- and two-phonon states with substantial weight.    
\section{Summary}
To conclude, investigating finite quantum structures 
coupled to vibronic degrees of freedom in the framework 
of a generalized Holstein-Hubbard Hamiltonian, we have demonstrated 
that interesting new physics, such as intrinsic (bi)polaron localization 
or phonon-assisted transmission, emerges when the energy scales 
set by external potentials, Coulomb and electron-phonon 
interactions  become comparable.
In this regime the interplay between the linear effects 
resulting from the barriers/cavities and the nonlinearity
inherent in a discrete interacting electron-phonon system
is of major importance. A general understanding of vibrational effects 
in (molecular) quantum transport, however, is still far off.   
Our objects in view will be to study (i) how polaronic quasiparticles 
time evolve when passing through phonon-coupled nanoscale structures 
and (ii) how finite temperature (heating) affects the balance between coherent 
and incoherent transport mechanisms.   
\section*{Acknowledgments}
The authors would like to thank A. Alvermann, G. Schubert, 
and S. A. Trugman for useful discussions. 
This work was supported by DFG through SFB 512 and Grant
No. 436 TSE 113/33, KONWIHR Bavaria, Academy of Sciences of the 
Czech Republic, and U.S. DOE. Numerical calculations were performed 
at LRZ Munich. H.F. and G.W. acknowledge hospitality at the 
Los Alamos National Laboratory and the Institute of Physics AS CR.


\begin{thebibliography}{10}
\bibitem{phononfrequencycomment}
To simplify the considerations we assume the phonon
frequencies to be the same at every site, ignoring that
a realistic system may be made of different materials.
It is commonly accepted, however, that the polaron and bipolaron 
effects in quantum dots are only weakly influenced be the details
of the optical phonon spectra.
\bibitem{estimate}
As a very crude estimate $(\varepsilon_{p}+\varepsilon_{p,4})$
has to overcome (i) the on-site potential $\Delta_4=4t$ 
and (ii) counterbalance the kinetic energy loss $2t$, 
i.e., taking  $\varepsilon_{p}=0.5t$ into account, 
we obtain $\varepsilon_{p,4}^c \sim 5.5t$ in surprisingly 
good agreement with the exact result from Fig.~\ref{Fig2}.
\end{thebibliography}
\end{document}